\documentclass[aps,
prd,
preprint,
longbibliography,
nofootinbib
]{revtex4-1}

\usepackage[dvipdfmx]
{graphicx}
\usepackage{amssymb}
\usepackage{amsmath}
\usepackage{braket}
\usepackage{listings}
\usepackage{cases}
\usepackage{comment}
\usepackage[colorlinks=true,
urlcolor=blue,
anchorcolor=blue,
citecolor=blue,
filecolor=blue,
linkcolor=blue,
menucolor=blue,
linktocpage=true,
pdfproducer=medialab,
pdfa=true
,dvipdfmx
]{hyperref}

\def\slashchar#1{\setbox0=\hbox{$#1$} 
\dimen0=\wd0 
\setbox1=\hbox{/} \dimen1=\wd1 
\ifdim\dimen0>\dimen1 
\rlap{\hbox to \dimen0{\hfil/\hfil}} 
#1 
\else 
\rlap{\hbox to \dimen1{\hfil$#1$\hfil}} 
/ 
\fi}

\begin{document}

\title{Four-body correlation embedded in antisymmetrized geminal power wave function}
\date{\today}
\author{Airi Kawasaki}
\email{a\_kawasaki@issp.u-tokyo.ac.jp}
\author{Osamu Sugino}
\email{sugino@issp.u-tokyo.ac.jp}
\affiliation{Institute for Solid State Physics, The University of Tokyo, 5-1-5 Kashiwanoha, Kashiwa, Chiba, 277-8581, Japan}

\begin{abstract}
We extend the Coleman's antisymmetrized geminal power (AGP) to develop a wave function theory that can incorporate up to four-body correlation in a region of strong correlation. To facilitate the variational determination of the wave function, the total energy is rewritten in terms of the traces of geminals. This novel trace formula is applied to a simple model system consisting of one dimensional Hubbard ring with a site of strong correlation. Our scheme significantly improves the result obtained by the AGP-CI scheme of Uemura {\it et al.}\ and also achieves more efficient compression of the degrees of freedom of the wave function. We regard the result as a step toward a first-principles wave function theory for a strongly correlated point defect or adsorbate embedded in an AGP-based mean-field medium.
\end{abstract}


\maketitle

\section{Introduction}

Determination of the ground state of a matter has been one of the most important problems common to condensed matter physics, quantum chemistry, and nuclear physics. The determination of the wave function is hindered by the extremely large degrees of freedom of Hilbert space even when the number of fermions $N$ is not so large. Recently, however, there was a progress in describing the wave function using significantly reduced degrees of freedom. This was achieved by density matrix renormalization group (DMRG) \cite{1992PhRvL..69.2863W}, tensor network renormalization group \cite{2015PhRvL.115t0401E}, and related wave function theories. These theories may be regarded as a kind of data compression, which is important not only in searching larger Hilbert space for better accuracy but is also in characterizing wave function by utilizing the compressed degrees of freedom. In this context, we tried in the past to compress the degrees of freedom by decomposing the antisymmetric tensor of degree $N$ appearing in the configuration interaction (CI) coefficient into tensors of lower degrees \cite{2012PhRvL.109y3001U,2015PhRvA..91f2504U}. Therein a classical canonical decomposition \cite{hitchcock-sum-1927} was applied to the symmetric part of the CI coefficient and, by optimizing the parameters, the wave function was compactly described taking only several terms for small molecules as well as for small Hamiltonian clusters. Each term in the series, called the symmetric tensor decomposition (STD) series, was represented by using the antisymmetrized geminal power (AGP) \cite{1965JMP.....6.1425C},
\begin{eqnarray}
\ket{F} &=& \left( \sum_{ab}F_{ab}c^{\dag}_{a}c^{\dag}_{b}\right) ^{\frac{N}{2}} \ket{0} \label{eq1}
\end{eqnarray}
where $F$ is an antisymmetric matrix of dimension $M$, with $M$ being the number of basis functions. In this paper, we consider systems of even number of particles only. Total number of degrees of freedom is $KM(M-1)/2$ with $K$ being the number of  terms in the STD series, which is much smaller than the number required in the conventional CI where the wave function is expanded by the Slater determinants. This method, called the extended STD (ESTD) or might be better called as AGP-CI, is based on the CI expansion method, so that $K$ is expected to increase rapidly with $N$. Therefore, the degrees of freedom should be reduced further to study larger systems.

The AGP wave function (Eq.\ (\ref{eq1})) is written in real space as
\begin{eqnarray}
\psi^{AGP}(r_{1},r_{2},\cdots ) = \hat{A}[g(r_{1},r_{2})g(r_{3},r_{4})\cdots] \label{eq2}
\end{eqnarray}
where $\hat{A}$ is the operator to fully antisymmetrize the geminals, $g(r_{1}, r_{2})$, which are defined using one-body orbitals $\varphi_{i}(r)$ and the Levi-Civita $\epsilon_{ij}$ as
\begin{eqnarray}
g(r_{1}, r_{2}) = \sum_{ij}\epsilon_{ij}\varphi_{i}(r_{1})\varphi_{j}(r_{2}) .
\end{eqnarray}

The AGP wave function describes correlation within the geminal correctly, but does not describe the correlation between geminals indicating that AGP is a mean-field theory for geminals. The correlation between two geminals can be described by introducing four-body extension of the geminal, $g_{4}(r_{1}, r_{2}, r_{3}, r_{4})$, and antisymmetrize the product as did in Eq.\ (\ref{eq2}), and the correlation between $g_{4}$ can be described by the antisymmetrized product of $g_{8}$, and so on. Such extension of AGP is our basic idea behind, and our aim is to investigate effectiveness of introducing this four-body correlation, instead of a Jastrow factor \cite{2003JChPh.119.6500C,2015JChPh.143x4112S}, into the AGP wave function. To simplify the problem, we try in this study to describe the four-body correlation within a small region (A) where the on-site Coulomb interaction is set stronger than in the rest of the system.

Specifically, we will use a trial wave function by multiplying the correlation factor of the form
\begin{eqnarray}
\exp\left(\sum_{ij}^{\text{A}} \sum_{pq}^{\text{all}}G_{ijpq}c^{\dag}_{i}c^{\dag}_{j}c^{\dag}_{p}c^{\dag}_{q}\right)
\end{eqnarray}
with the AGP wave function, Eq.\ (\ref{eq1}), using the antisymmetric tensor of degree four ($G_{ijpq}$). We will exemplify the effectiveness of the approach by using a one-dimensional Hubbard model having specifically large $U$ on a single site; namely we will use a simple impurity problem for the test. Because of the single site used for the region A, the number of additional degrees of freedom for our trial wave function is $M(M-1)/2$, which is the same as that for the APG wave function.

To calculate the total energy variationally, a simple-minded formulation can lead the computation impractically demanding because of large number of terms thereby generated. We thus need to advance our previous technique even though we are using a minimal model. To approach this problem, we utilize algorithms developed in the field of nuclear physics \cite{onishi1966generator}, \cite{2012PhLB..715..219M} and an original trace formalism to develop a systematic formula for the expectation value of the Hamiltonian. We will show how the novel formula works effectively in determining the total energy variationally with the help of the standard conjugate gradient minimization technique.\\

This paper is organized as follows. 
In section~\ref{Formulation}, we show a formalism to calculate the Hamiltonian matrix element using the AGP basis.
Then we show the calculated results and compare them with AGP-CI in section~\ref{Result} followed by conclusion in section~\ref{conclusions}.

\section{Formulation}
\label{Formulation}

\subsection{Density matrices}

The advantage of using AGP as the basis is that the matrix elements of the Hamiltonian as well as the overlap are given analytical form. Using the formula of Onishi and Yoshida \cite{onishi1966generator}, the overlap of an $N$ particle system can be given by the $N$-th order coefficient of the polynomial of an auxiliary variable $t$ as,
\begin{eqnarray}
\left.\braket{tF^{\lambda}|tF^{\mu}}\right| _{t^{N}} &=& \left.\exp\left(  \frac{1}{2}\mathrm{tr}\left[  \ln(1+F^{\mu}F^{\lambda\dag}t^{2})\right] \right)\right| _{t^{N}}\nonumber\\
&=& \left.\mathrm{pf}(1+F^{\mu}F^{\lambda\dag}t^{2})\right| _{t^{N}} , \label{norm}
\end{eqnarray}
where pf is Fredholm Pfaffian and $F^{\lambda\dag}$ is the Hermitian conjugate of $F^{\lambda}$. Matrix element of the one-body term of the Hamiltonian can be obtained from the first-order density matrix
\begin{eqnarray}
&&\left.\bra{tF^{\lambda}}c^{\dag}_{a}c_{b}\ket{tF^{\mu}} \right| _{t^{N}}\nonumber\\
&&= \left[\frac{F^{\mu}F^{\lambda\dag}t^{2}}{1+F^{\mu}F^{\lambda\dag}t^{2}}\right]_{ba} \left.\mathrm{pf}(1+F^{\mu}F^{\lambda\dag}t^{2})\right| _{t^{N}} .
\end{eqnarray}
The two-body term can be obtained from the second-order density matrix
\begin{eqnarray}
&&\left.\bra{tF^{\lambda}}c^{\dag}_{p}c^{\dag}_{q}c_{s}c_{r}\ket{tF^{\mu}} \right| _{t^{N}}\nonumber\\
&=& \left( \left[\frac{F^{\mu}F^{\lambda\dag}t^{2}}{1+F^{\mu}F^{\lambda\dag}t^{2}}\right]_{rp}\left[\frac{F^{\mu}F^{\lambda\dag}t^{2}}{1+F^{\mu}F^{\lambda\dag}t^{2}}\right]_{sq} \right.\nonumber \\
&& -\left[\frac{F^{\mu}F^{\lambda\dag}t^{2}}{1+F^{\mu}F^{\lambda\dag}t^{2}}\right]_{rq}\left[\frac{F^{\mu}F^{\lambda\dag}t^{2}}{1+F^{\mu}F^{\lambda\dag}t^{2}}\right]_{sp} \nonumber\\
&& \left. +\left[\frac{t}{1+F^{\mu}F^{\lambda\dag}t^{2}}F^{\mu}\right]_{rs}\left[F^{\lambda\dag}\frac{t}{1+F^{\mu}F^{\lambda\dag}t^{2}}\right]_{qp} \right)\nonumber\\
&&\times\left.\mathrm{pf}(1+F^{\mu}F^{\lambda\dag}t^{2})\right| _{t^{N}} . \label{2}
\end{eqnarray}
These formulae can be derived from the commutation relation
\begin{eqnarray}
\left[  c_{\alpha},\exp\left(  \sum_{\alpha^{\prime}\beta^{\prime}%
}F_{\alpha^{\prime}\beta^{\prime}}c_{\alpha^{\prime}}^{\dag}c_{\beta^{\prime}%
}^{\dag}t\right)  \right]  \nonumber\\
=\sum_{\gamma}tF_{\alpha\gamma}c_{\gamma}^{\dag}%
\exp\left(  \sum_{\alpha^{\prime}\beta^{\prime}}F_{\alpha^{\prime
}\beta^{\prime}}c_{\alpha^{\prime}}^{\dag}c_{\beta^{\prime}}^{\dag}t\right)
\label{OYc}%
\end{eqnarray}
which yields
\begin{eqnarray}
c_{\alpha} \ket{tF} =  \sum_{\gamma}tF_{\alpha\gamma}c^{\dag}_{\gamma} \ket{tF},
\end{eqnarray}
and the fact that
\begin{eqnarray}
\bra{tf}c^{\dag}_{\alpha}c_{\beta}\ket{tF} & = & \bra{tf}c^{\dag}_{\alpha}\sum_{\gamma}tF_{\beta\gamma}c^{\dag}_{\gamma}\ket{tF} \nonumber\\
& = & \sum_{\gamma}F_{\beta\gamma}\frac{\partial}{\partial F_{\alpha\gamma}}\braket{tf|tF},
\end{eqnarray}
which is valid to general antisymmetric matrices $F$ and $f$.

In the extended symmetric tensor decomposition (ESTD) theory, or the AGP-CI theory, of Uemura {\it et al.}, the above formulae were used to vary the germinal matrices $F^\lambda$ with $\lambda = 1\cdots K$ to obtain the total energy.

\subsection{Correlation factor and trace formula}

We introduce a correlation factor to take into account the correlation of geminals. Our trial wave function is chosen as
\begin{eqnarray}
\ket{\Psi} = \left.\exp\left[\sum_{ij}^{\text{A}}\sum_{p q}^{\text{all}}G_{i j p q}c^{\dag}_{i}c^{\dag}_{j}c^{\dag}_{p}c^{\dag}_{q}t^{2}\right] \Ket{tF}\right|_{t^{N}} \label{hado},
\end{eqnarray}
to include the four-body correlation in the region A where the correlation is strong. This can be understood, for example, from the fact that the antisymmetric tensor of degree four $G$ can be decomposed into a product of antisymmetric matrices $g$ as
\begin{eqnarray}
G_{i j p q} = \sum_{\nu}g^{\nu}_{ij}g^{\nu}_{pq}+g^{\nu}_{ip}g^{\nu}_{qj}+g^{\nu}_{iq}g^{\nu}_{jp} \label{g}
\end{eqnarray}
with each term in the right-hand side of Eq.\ (\ref{g}) representing correlation of two geminals in the region A. Note that we neglect the correlation beyond the two geminals, say, ``four-body extension of the geminal''.

The corresponding overlap and Hamiltonian matrix elements can be obtained by expanding the correlation factor into series and by applying the above formulae Eqs.(\ref{norm})-(\ref{2}). General terms in the series involve,
\begin{eqnarray}
\bra{tF^{\lambda}}c_{a_1}c_{a_2}\cdots c_{a_n}c^{\dag}_{b_1}c^{\dag}_{b_2}\cdots c^{\dag}_{b_m}\ket{tF^{\mu}}, \label{ff}
\end{eqnarray}
which can be rewritten using the formula of Mizusaki and Oi \cite{2012PhLB..715..219M} in terms of the Pfaffian
\begin{eqnarray}
&&\mathrm{Pf}(Z)_{i_{2N},i_{2N-1},\cdots , i_{4},i_{3},i_{2},i_{1}} \nonumber\\
&&\equiv \sum_{\sigma \in S_{2N}}\textrm{sgn}(\sigma) Z_{\sigma (i_{1})\sigma(i_{2})}Z_{\sigma (i_{3})\sigma(i_{4})}\cdots Z_{\sigma (i_{2N-1})\sigma(i_{2N})} \nonumber\\
\label{pf}
\end{eqnarray}
of the inverse of a $2M\times 2M$ antisymmetric matrix
\begin{eqnarray}
X = \left(
\begin{array}{ cc}
    tF^{\mu}  &   1 \\
    -1  &  tF^{\lambda\dagger}
\end{array}\right),
\end{eqnarray}
such that
\begin{eqnarray}
&&\bra{tF^{\lambda}}c_{a_1}c_{a_2}\cdots c_{a_n}c^{\dag}_{b_1}c^{\dag}_{b_2}\cdots c^{\dag}_{b_m}\ket{tF^{\mu}} \nonumber\\
&&= \mathrm{Pf}(X^{-1})_{b_{m},\cdots , b_{1},a_{n}+M,\cdots ,a_{1}+M} \braket{tF^{\lambda}|tF^{\mu}}\label{MO}
\end{eqnarray}
with
\begin{eqnarray}
X^{-1} 
= \left(
\begin{array}{ cc}
    F^{\lambda\dagger}t\frac{1}{1+F^{\mu}F^{\lambda\dag}t^{2}}  &   \frac{1}{1+F^{\lambda\dag}F^{\mu}t^{2}}  \\
    -\frac{1}{1+F^{\mu}F^{\lambda\dag}t^{2}}  &   \frac{1}{1+F^{\mu}F^{\lambda\dag}t^{2}}F^{\mu}t
\end{array}\right) \label{eq:yon1} 
\equiv \left(
\begin{array}{ cc}
    Z^{1}  &   Z^{3} \\
    Z^{4}  &   Z^{2}
\end{array}\right). \label{z}
\end{eqnarray}
Here $M$ is the dimension of the antisymmetric matrix $F$, and $S_{2N}$ is the permutation group of element $2N$.  With Eq.\ (\ref{MO}), the overlap and the Hamiltonian matrix elements can be written by terms of the form
\begin{eqnarray}
&&P_N (A^{(1)}, A^{(2)}, \cdots , A^{(N)}) \nonumber\\
&\equiv& \sum_{i_1 , i_2 , \cdots } A^{(1)}_{i_1, i_2} A^{(2)}_{i_3, i_4} \cdots A^{(N)}_{i_{2N-1}, i_{2N}} \nonumber\\
&\times& \sum_{\sigma\in S_{2N}} \textrm{sgn}(\sigma) X^{-1}_{\sigma (i_1 ) \sigma (i_2 )} X^{-1}_{\sigma (i_3 ) \sigma (i_4 )} \cdots X^{-1}_{\sigma (i_{2N-1} ) \sigma (i_{2N} )} , \nonumber \\
\label{pn}
\end{eqnarray}
after applying the antisymmetric matrix decomposition like Eq.\ (\ref{g}) and extending the matrices to $2M\times 2M$ as
\begin{eqnarray}
g^{\nu}\rightarrow\left(
\begin{array}{ cc}
    g^{\nu}  &   0 \\
    0  &   0
\end{array}\right) \text{ and }
g^{\nu\dag}\rightarrow\left(
\begin{array}{ cc}
    0  &   0 \\
    0  &   g^{\nu\dag}
\end{array}\right).
\end{eqnarray}
Since each index appears twice in this expression, Eq.\ (\ref{pn}) should be rewritten as a product-sum of traces of the form
\begin{eqnarray}
&&c_{1}\text{tr}[A^{(1)}X^{-1}]\text{tr}[A^{(2)}X^{-1}]\text{tr}[A^{(3)}X^{-1}]\cdots \nonumber\\
&+& c_{2}\text{tr}[A^{(1)}X^{-1}]\text{tr}[A^{(2)}X^{-1}A^{(3)}X^{-1}]\cdots \nonumber\\
&+& \cdots . 
\end{eqnarray}
By investigating the terms, we have found that the coefficient $c_i$ is simply given as
\begin{eqnarray}
c_i = \frac{2^{N}}{(-2)^{k_{1}+k_{2}+\cdots}}.
\end{eqnarray}
when the number of traces of length one, such as $\text{tr}[A^{(1)}X^{-1}]$, contained in the term is $k_1$, the number of traces of length two, such as $\text{tr}[A^{(1)}X^{-1}A^{(2)}X^{-1}]$, is $k_2$ and so on. Therefore, the total number of traces ($N_{\text{tr}}$) is important: Eq.\ (\ref{pn}) can be rewritten as a sum of all possible combinations of the product of the traces with the weight for the summation taken to be $\frac{2^N}{(-2)^{N_{\text{tr}}}}$. This fact drastically facilitates the numerical evaluation.

Although the resulting formulae, which we call trace formula, are greatly simplified, they become more and more complex as the size of the region A increases. So, let us restrict our study to the case where the region A consists of single site of index $i$. In other words, we will study the problem of a single impurity in a mean-field type AGP medium. We further restrict the degrees of freedom for the $G$ in Eq.\ (\ref{g}) such that $G_{klpq}$ has a value only when two of the indices $k$ and $l$ are in the region A, namely belonging to the site $i$, and others $p$ and $q$ are not. By this, the exponent of Eq.\ (\ref{hado}) becomes
\begin{eqnarray}
\sum_{kl}^{\in [i,\overline{i}]}g_{kl}c^{\dag}_{k}c^{\dag}_{l}\sum_{pq}^{{\in\hspace{-.45em}/} [i,\overline{i}]}G^{kl}_{pq}c^{\dag}_{p}c^{\dag}_{q}t^2 , \label{yo}
\end{eqnarray}
where $i$ and $\overline{i}$ correspond to the state with spin-up and spin-down, respectively, and $g_{kl}$ is taken to be $\pm 1/2$ only when $\{ k,l\} =\{ i,\overline{i}\}$ and $\{\overline{i},i\}$, respectively. $G_{pq}^{kl}$ is antisymmetric with respect to the subscripts and depends parametrically on the combination indices $k$ and $l$; since there is only one combination of $i$ and $\overline{i}$, we will omit the superscript. Under this simplification, the overlap is given by the $N$-th order coefficient of
\begin{eqnarray}
\bra{tF}\left( 1+\sum_{klpq}g_{kl}G_{pq}c^{\dag}_{k}c^{\dag}_{l}c^{\dag}_{p}c^{\dag}_{q}t^2 + \sum_{mnrs}g_{mn}^{\dag}G_{rs}^{\dag}c_{s}c_{r}c_{n}c_{m}t^2 \right. \nonumber\\
\left.+ \sum_{klmnpqrs}g_{mn}^{\dag}G_{rs}^{\dag}g_{kl}G_{pq}c_{s}c_{r}c_{n}c_{m}c^{\dag}_{k}c^{\dag}_{l}c^{\dag}_{p}c^{\dag}_{q}t^4  \right) \ket{tF} \nonumber\\
\label{tenkai}
\end{eqnarray}
where indices, $k,l,n$ and $m$, correspond to the site $i$, while other indices are not. Then, contribution from the fourth term in the parenthesis of Eq.\ (\ref{tenkai}), for example, is rewritten as
\begin{eqnarray}
& & \sum_{klmnpqrs}g^{\dag}_{mn}G^{\dag}_{pq}g_{kl}G_{rs}\mathrm{Pf}(X^{-1})_{k, l, p, q, r+M, s+M, m+M, n+M} \nonumber\\
 = && (g^{\dag})(g)(G^{\dag})(G) \nonumber\\
 &-&2[(g^{\dag}g)(G^{\dag})(G) + (g^{\dag}G^{\dag})(g)(G) + (g^{\dag}G)(g)(G^{\dag}) \nonumber\\
&&+ (gG^{\dag})(g^{\dag})(G) + (gG)(g^{\dag})(G^{\dag}) + (G^{\dag}G)(g^{\dag})(g) ] \nonumber\\
&  +&4[(g^{\dag}gG^{\dag})(G) + (g^{\dag}gG)(G^{\dag}) + (g^{\dag}G^{\dag}G)(g) + (gG^{\dag}G)(g^{\dag}) \nonumber\\
&&+ (g^{\dag}G^{\dag}g)(G) + (g^{\dag}Gg)(G^{\dag}) + (g^{\dag}GG^{\dag})(g) + (gGG^{\dag})(g^{\dag}) \nonumber\\
& & + (g^{\dag}g)(G^{\dag}G) + (g^{\dag}G^{\dag})(gG) + (g^{\dag}G)(gG^{\dag}) ] \nonumber\\
&  -&8[(g^{\dag}gG^{\dag}G) + (g^{\dag}gGG^{\dag}) + (g^{\dag}G^{\dag}gG) \nonumber\\
&&+ (g^{\dag}G^{\dag}Gg) + (g^{\dag}GgG^{\dag}) + (g^{\dag}GG^{\dag}g) ]
\end{eqnarray}
where we have used the abbreviations such as $(g^{\dag}gG^{\dag}G) = \text{tr}[g^{\dag}Z^{4}gZ^{3}G^{\dag}Z^{4}GZ^{3}]$ with $Z^i$ being the quantity defined in Eq.\ (\ref{z}). Other terms of Eq.\ (\ref{tenkai}) can be similarly rewritten. Note that the same result can be alternatively obtained using the recursion formula of $P_N$
\begin{eqnarray}
&&P_N (A^{(1)}, A^{(2)}, \cdots , A^{(N)}) \nonumber\\
&=&-(A^{(1)}) P_{N-1} (A^{(2)}, \cdots , A^{(N)}) \nonumber\\
&&+ 2\sum_{k=2}^N P_{N-1} (A^{(2)}, \cdots , A^{(k-1)} , \nonumber\\
&& \;\;\;\;\;\;\;\;\;\;\;\;\;\;\;\;\;\;\;\;\;A^{(1)}X^{-1}A^{(k)}, A^{(k+1)},\cdots ,  A^{(N)}), \nonumber\\
\label{re}
\end{eqnarray}
which can be derived from the known expansion formula for Pfaffian, 
\begin{eqnarray}
&&\mathrm{Pf}(Z)_{i_{N},\cdots ,i_{1}} \nonumber\\
&&= \sum_k (-1)^{k}Z_{i_N , i_k}\mathrm{Pf}(Z)_{\hat{i}_{N},i_{N-1},\cdots , \hat{i}_{k},\cdots,i_{1}}, \label{rec}
\end{eqnarray}
where $\hat{i}_{k}$ means to omit the index $i_k$.

We can derive, in a similar way, formula for the Hamiltonian matrix elements and its derivative with respect to $F$ and $G$, which will be detailed in Appendix \ref{app:decay}. Note that directly differentiating by $F$ is too tedious: Instead, it is simpler to use the relation
\begin{eqnarray}
\frac{1}{t}\frac{\partial}{\partial F_{cd}}\braket{tf|tF} = \bra{tf}c^{\dag}_{c}c^{\dag}_{d}\ket{tF}\label{bibun}
\end{eqnarray}
and then apply the trace formula. Using the above algorithms, it is straightforward to give a formula for the total energy and its parameter derivatives, which can be used to determine the total energy fully variationally.

\section{Result}
\label{Result}

\begin{figure}[htbp]
  \begin{center}
    \begin{tabular}{c}
    \hspace*{1.5cm} 
          \includegraphics[clip, width=5.0cm]{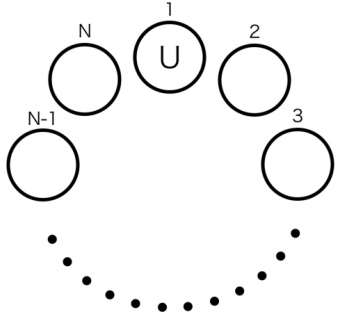}
          \hspace{1.6cm} 
     \end{tabular}
    \caption{One-dimensional Hubbard model with periodic boundary condition. Only on the first site nonzero value $U$ is assigned as the on-site Coulomb interaction.}
    \label{fig:zu1}
  \end{center}
\end{figure}

We use a periodic one-dimensional tight-binding model of length $N$ with nearest neighbor transfer only ($t_0 = -1$), or the Hubbard ring, and assign zero for the on-site energy. The on-site Coulomb interaction is non-zero only at the first site as shown in Fig.\ \ref{fig:zu1}. The total energy is variationally determined using the trial wave function (Eq.\ (\ref{hado}), (\ref{yo})), which is referred to as AGP4, and also using the AGP-CI trial wavefunction
\begin{eqnarray}
\ket{\Psi} = \sum_{\lambda = 1}^{K}\ket{F^{\lambda}} ,
\end{eqnarray}
which is referred to as AGP-CI($K$) for $K=1, 2$, and $3$. AGP-CI($1$) will be sometimes denoted as AGP.

\begin{figure}[htbp]
  \begin{center}
    \begin{tabular}{c}
      \begin{minipage}{0.33\hsize}
         \hspace*{-3.5cm} 
          \includegraphics[clip, width=9.0cm]{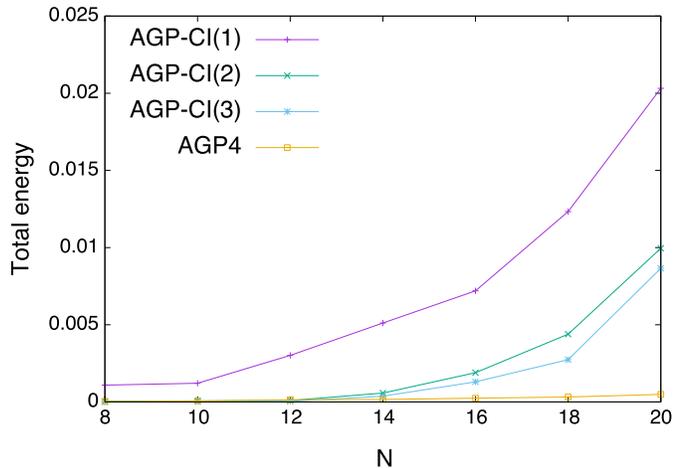}
          \hspace{1.6cm} 
      \end{minipage}
     \end{tabular}
    \caption{(Color online) Total energy of the Hubbard ring plotted against the length $N$. The energy is referred to that obtained by the exact diagonalization.}
    \label{fig:zu2}
  \end{center}
\end{figure}

Fig.\ \ref{fig:zu2} shows a comparison of the total energy obtained for the Hubbard ring of length $8$ to $20$ ($N=8\cdots20$) and for the on-site Coulomb interaction $10$ ($U=10$). All the total-energies are referred to that obtained by applying the exact diagonalization \cite{caesar} to the ring of the same length. 

For AGP-CI($K$), we can see a super-linear deviation from the exact diagonalization, and the deviation is appreciable, in the energy scale of Fig.\ \ref{fig:zu2}, already at $N=8$ for $K=1$ and at $N=14$ for $K=2$ and $3$. By contrast, deviation remains small for the AGP4 even at $N=20$, where the value is $0.00048$. Considering that the number of parameters is the same for AGP-CI($2$) and AGP$4$, the wave function is much more efficiently compressed by AGP$4$.

In addition to the superior result for the total energy, AGP$4$ is numerically more attractive. Indeed, for AGP-CI's, we need more than $50$ trial runs from different initial parameters to obtain a reliable value for the total energy because the calculation with the standard conjugate gradient minimization is very easily trapped at a local minimum. See Appendix \ref{app:decay_mode} for more detail. Note in addition that Ref.\ \cite{2015PhRvA..91f2504U} reported that information loss easily occurs although that is not so important in the present case. Whereas, we need only $1$ trial run for AGP$4$. We conjecture the large number of local minima found for AGP-CI will be due to the known instability problem that the canonical decomposition has \cite{2009SIAMR..51..455K}.

\begin{figure}[htbp]
  \begin{center}
    \begin{tabular}{c}
      \begin{minipage}{0.33\hsize}
         \hspace*{-3.5cm} 
          \includegraphics[clip, width=9.0cm]{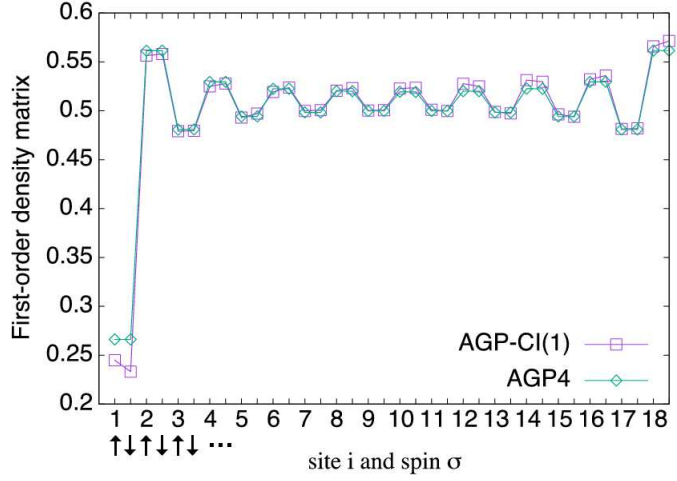}
          \hspace{1.6cm} 
      \end{minipage}
     \end{tabular}
    \caption{(Color online) First-order density matrix ($\rho_{i\sigma i\sigma}^{(1)}$) of the Hubbard ring of length $N=18$. The value for up-spin and that for down spin are arranged alternately.}
    \label{fig:zu3}
  \end{center}
\end{figure}

Fig.\ \ref{fig:zu3} shows density matrices (DM's) of the Hubbard ring of length $18$ obtained by AGP and AGP$4$. The value of $U$ is taken to be $10$. The diagonal element of the first-order DM ($\rho_{i\sigma i\sigma}^{(1)}$) shows that the difference between the two calculations is most strikingly found at the first site with nonzero $U$. For AGP, the density for the up-spin is different from that of the down-spin, indicating a symmetry-broken solution is obtained, while the spin symmetry is maintained for AGP$4$ even though the symmetry is not intentionally restricted. The value of the first-order DM is larger for AGP$4$ at the first site, indicating that there is more probability finding a particle at this site for AGP$4$.

\begin{figure}[htbp]
  \begin{center}
    \begin{tabular}{c}
      \begin{minipage}{0.33\hsize}
        \begin{center}
         \hspace*{-3.5cm} 
          \includegraphics[clip, width=9.0cm]{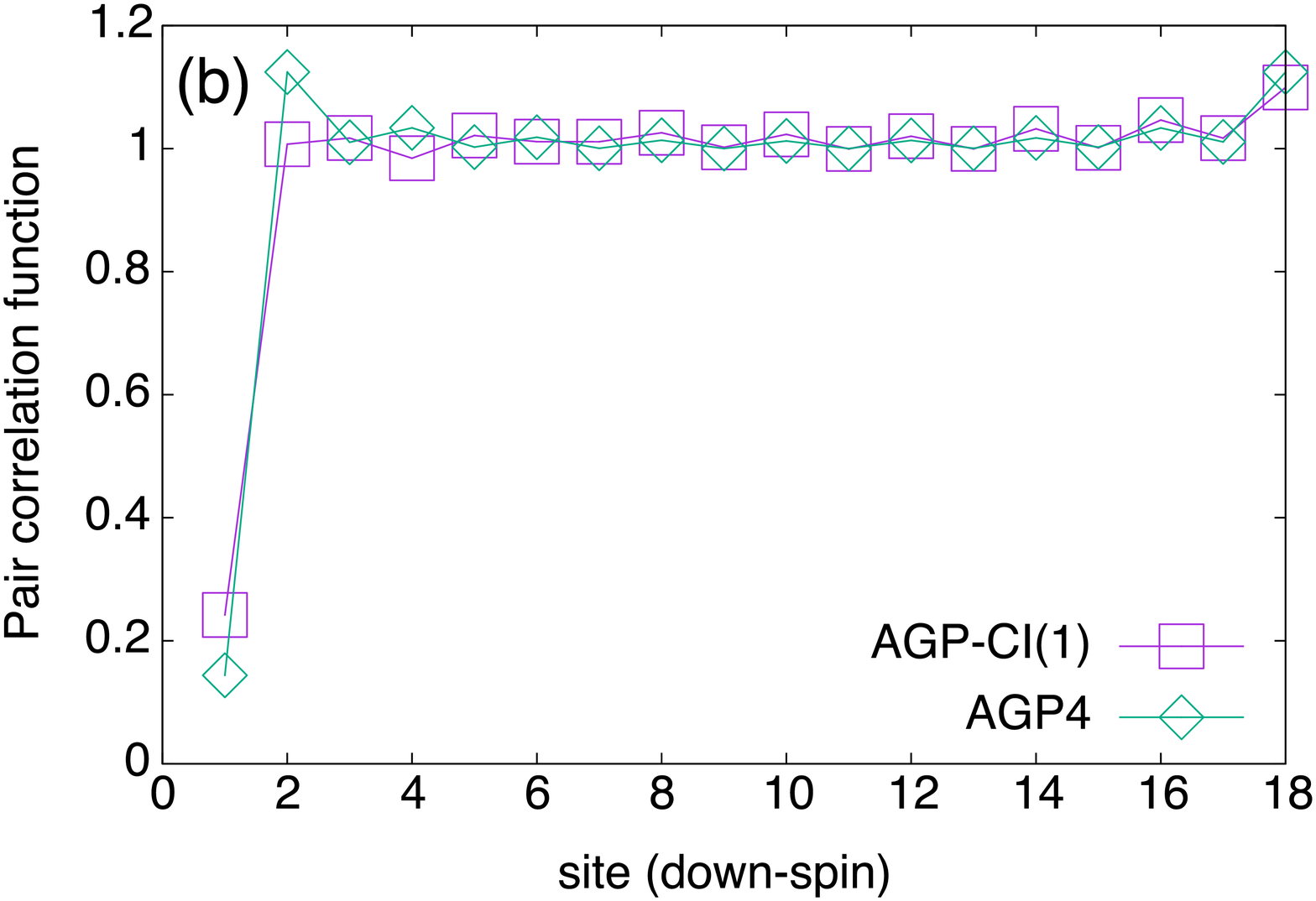}
          \hspace{1.6cm} 
        \end{center}
      \end{minipage}
       \end{tabular}
    \caption{(Color online) Pair correlation function ($\rho_{1\uparrow i\sigma}^{(2)}/2\rho_{1\uparrow 1\uparrow}^{(1)}\rho_{i\sigma i\sigma}^{(1)} $) of the Hubbard ring of length $N=18$. (a) and (b) correspond, respectively, to $\sigma = \uparrow$ and $\sigma = \downarrow$.}
    \label{fig:zu4}
  \end{center}
\end{figure}

For the pair correlation function between the particle with up-spin at the first site and that with spin $\sigma$ at $i$-th site ($f_{i\sigma}\equiv \rho_{1\uparrow i\sigma}^{(2)}/2\rho_{1\uparrow 1\uparrow}^{(1)}\rho_{i\sigma i\sigma}^{(1)} $), the difference between the two calculations is most striking for the down-spin at the first site and its nearest neighbor sites. Fig.\ \ref{fig:zu4} shows that $f_{1\downarrow}$ of AGP4 is smaller than that of AGP indicating that electrons are repelled more strongly at the first site. For AGP4, we can regard the particle is repelled to the nearest neighbor sites when considering that $f_{1\downarrow}$ and $f_{18\downarrow}$ are larger than the value at different site ($N=3\cdots 17$). For AGP, on the other hand, the particle is repelled to one of the nearest neighbor sites, breaking thereby the symmetry as seen in the above.

 \begin{figure}[htbp]
  \begin{center}
    \begin{tabular}{c}
      \begin{minipage}{0.33\hsize}
        \begin{center}
         \hspace*{-3.5cm} 
          \includegraphics[clip, width=9.0cm]{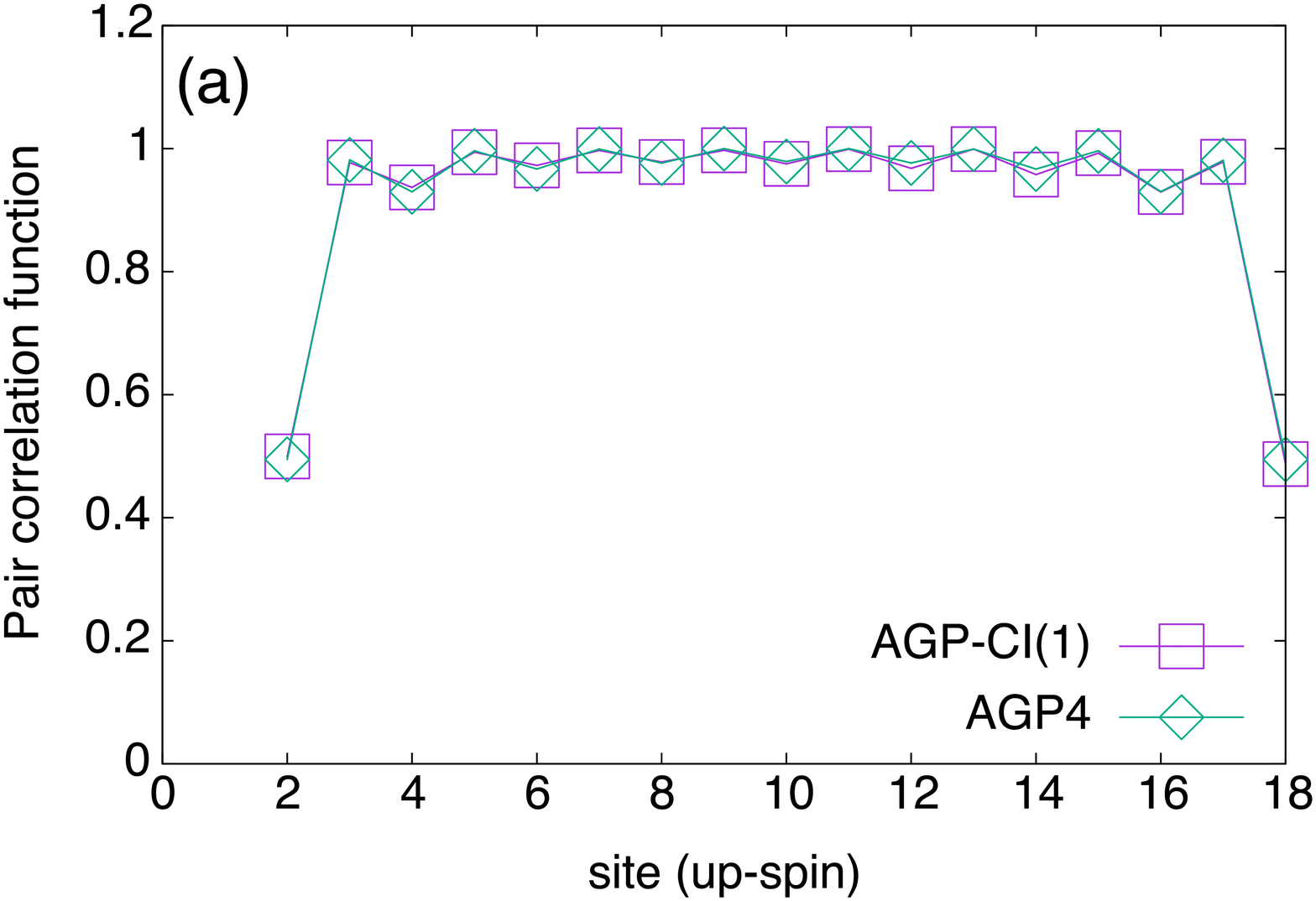}
          \hspace{1.6cm} 
        \end{center}
      \end{minipage}
       \end{tabular}
  \end{center}
\end{figure}

\section{SUMMARY AND CONCLUSION}
\label{conclusions}

We extended the AGP-CI scheme of Uemura {\it et al.} to incorporate up to four-body correlation in a region of interest. For this purpose, we constructed a trial function by multiplying a four-body correlation factor with the AGP wave function. To facilitate the variational determination of the total energy using the standard conjugate gradient minimization method, we greatly simplified the formula for the Hamiltonian matrix elements by rewriting them in terms of traces of the geminals. The trace formula was used to calculate a Hubbard ring having a site of strong correlation. The calculated total energy showed a significant improvement over AGP-CI. In addition, the new scheme showed enhanced numerical stability and robustness against the symmetry breaking. We also found better efficiency in the data compression of the wave function. Although we demonstrated using only the Hubbard ring, generalization to molecular systems is straightforward albeit being more complicated. We regard the result as an important step toward an establishment of first-principles wave function theory for a strongly correlated point defect or adsorbate embedded in an AGP-based mean-field medium.

\begin{acknowledgements}
The authors thank the Supercomputer Center, the Institute for Solid State Physics, the University of Tokyo for the use of the facilities. Exact diagonalization was done by using H$\Phi$ \cite{caesar}.
\end{acknowledgements}

\hspace{2cm}

\begin{widetext}
\appendix

\section{Details on the trace formula}
\label{app:decay}

Here we show details on the trace formula for the Hamiltonian matrix elements and their derivatives with respect to F and G. In our calculation, only the $\Gamma$ point is used and then $F$ and $G$ are taken to be real antisymmetric.

The overlap of the wave function $\braket{\Psi | \Psi }$ is given by
\begin{eqnarray}
\left( 1+ P_2 (\overline{g}, \overline{G})t^2 +  P_2 (\overline{g^{\dagger}}, \overline{G^{\dagger}})t^2 +  P_4 (\overline{g^{\dagger}}, \overline{G^{\dagger}}, \overline{g}, \overline{G})t^4\right) \braket{tF|tF} \label{a1}, 
\end{eqnarray}
using the matrix $G$, $G^{\dagger}$, $g$ and $g^{\dagger}$ extended to $2M\times 2M$ as
\begin{eqnarray}
\overline{G}=\left(
\begin{array}{ cc}
    G  &   0 \\
    0  &   0
\end{array}\right), 
\overline{G^{\dagger}}=\left(
\begin{array}{ cc}
    0  &   0 \\
    0  &   G^{\dagger}
\end{array}\right), 
\overline{g}=\left(
\begin{array}{ cc}
    g  &   0 \\
    0  &   0
\end{array}\right), 
\overline{g^{\dagger}}=\left(
\begin{array}{ cc}
    0  &   0 \\
    0  &   g^{\dagger}
\end{array}\right). \nonumber
\end{eqnarray}
The one-body term of the Hamiltonian
\begin{eqnarray}
\sum_{<i,j>}t_{0ij}\bra{\Psi} c^{\dag}_{i}c_{j}\ket{\Psi},  \nonumber  
\end{eqnarray}
\text{where $<i,j>$ means to take nearest neighbor site pair only}, is given by
\begin{eqnarray}
 \left( P_1 (\overline{t}) + P_3 (\overline{t}, \overline{g}, \overline{G})t^2 +  P_3 (\overline{t}, \overline{g^{\dagger}}, \overline{G^{\dagger}})t^2 +  P_5 (\overline{t}, \overline{g^{\dagger}}, \overline{G^{\dagger}}, \overline{g}, \overline{G})t^4\right) \braket{tF|tF} ,\label{a2}
\end{eqnarray} 
using the transfer matrix $t_{0ij}=\delta_{i+2,j}+\delta_{i,j+2}+\delta_{i,j+M-2}+\delta_{i+M-2,j}$ extended to $2M\times 2M$ as
\begin{eqnarray}
\overline{t}=\left(
\begin{array}{ cc}
    0  &   \frac{1}{2}t_0 \\
    -\frac{1}{2}t_0  &   0
\end{array}\right). \nonumber
\end{eqnarray}
Similarly, the two-body term of the Hamiltonian
\begin{eqnarray}
U\bra{\Psi} c^{\dag}_{1}c_{1}c^{\dag}_{\bar{1}}c_{\bar{1}}\ket{\Psi}   \nonumber
\end{eqnarray}
is given by 
\begin{eqnarray}
U \left( P_2 (\overline{u1}, \overline{u2}) + P_4 ( \overline{g}, \overline{G}, \overline{u1}, \overline{u2})t^2 +  P_4 (\overline{g^{\dagger}}, \overline{G^{\dagger}}, \overline{u1}, \overline{u2})t^2 +  P_6 (\overline{g^{\dagger}}, \overline{G^{\dagger}}, \overline{g}, \overline{G}, \overline{u1}, \overline{u2})t^4\right) \braket{tF|tF}, \nonumber\\
\label{a3}
\end{eqnarray}
using the matrix $u_{ij}=\frac{1}{2}(\delta_{i1}\delta_{j\bar{1}}-\delta_{j1}\delta_{i\bar{1}})$ extended to $2M\times 2M$ as
\begin{eqnarray}
\overline{u1}=\left(
\begin{array}{ cc}
    u  &   0 \\
    0  &   0
\end{array}\right), 
\overline{u2}=\left(
\begin{array}{ cc}
    0  &   0 \\
    0  &   u
\end{array}\right). \nonumber
\end{eqnarray}
Derivative of the overlap
\begin{eqnarray}
\frac{\partial}{\partial F_{ef}}\braket{\Psi | \Psi} \nonumber
\end{eqnarray}
is given by
\begin{eqnarray}
&& ( P_1 (\overline{d1})t + P_1 (\overline{d2})t +P_3 ( \overline{g}, \overline{G}, \overline{d1})t^3 +P_3 ( \overline{g}, \overline{G}, \overline{d2})t^3+  P_3(\overline{g^{\dagger}}, \overline{G^{\dagger}}, \overline{d1})t^3 +  P_3(\overline{g^{\dagger}}, \overline{G^{\dagger}}, \overline{d2})t^3\nonumber\\
&&+  P_5(\overline{g^{\dagger}}, \overline{G^{\dagger}}, \overline{g}, \overline{G}, \overline{d1})t^5+  P_5(\overline{g^{\dagger}}, \overline{G^{\dagger}}, \overline{g}, \overline{G}, \overline{d1})t^5) \braket{tF|tF} \label{a4},
\end{eqnarray} 
using the matrix $d_{ij}=\frac{1}{2}(\delta_{ie}\delta_{jf}-\delta_{if}\delta_{je})$ extended to $2M\times 2M$ as
\begin{eqnarray}
\overline{d1}=\left(
\begin{array}{ cc}
    d  &   0 \\
    0  &   0
\end{array}\right), 
\overline{d2}=\left(
\begin{array}{ cc}
    0  &   0 \\
    0  &   d
\end{array}\right). \nonumber
\end{eqnarray}
The derivative with respect to $G_{ef}$,
\begin{eqnarray}
\frac{\partial}{\partial G_{ef}} \braket{\Psi |\Psi} \nonumber
\end{eqnarray}
is given by
\begin{eqnarray}
 2( P_2 ( \overline{g}, \overline{d1})t^2 -P_2(\overline{g^{\dagger}}, \overline{d2})t^2 +  P_4(\overline{g^{\dagger}}, \overline{G^{\dagger}}, \overline{g}, \overline{d1})t^4-  P_4(\overline{g^{\dagger}}, \overline{g}, \overline{G}, \overline{d2})t^4) \braket{tF|tF} \label{a5}.
\end{eqnarray}
The derivatives of the Hamiltonian matrix elements are also obtained similarly:
\begin{eqnarray}
&&\frac{\partial}{\partial F_{ef}}\sum_{<i,j>}t_{0ij}\bra{\Psi} c^{\dag}_{i}c_{j}\ket{\Psi}\nonumber\\
&=&
( P_2 (\overline{t},\overline{d1})t + P_2 (\overline{t},\overline{d2})t \nonumber\\
&&+ P_4 (\overline{t}, \overline{g}, \overline{G}, \overline{d1})t^3+ P_4 (\overline{t}, \overline{g}, \overline{G}, \overline{d2})t^3 +  P_4 (\overline{t}, \overline{g^{\dagger}}, \overline{G^{\dagger}}, \overline{d1})t^3  +  P_4 (\overline{t}, \overline{g^{\dagger}}, \overline{G^{\dagger}}, \overline{d2})t^3 \nonumber\\
&&+  P_6 (\overline{t}, \overline{g^{\dagger}}, \overline{G^{\dagger}}, \overline{g}, \overline{G}, \overline{d1})t^5+  P_6 (\overline{t}, \overline{g^{\dagger}}, \overline{G^{\dagger}}, \overline{g}, \overline{G}, \overline{d2})t^5) \braket{tF|tF} 
\end{eqnarray}
\begin{eqnarray}
\frac{\partial}{\partial F_{ef}}U\bra{\Psi} c^{\dag}_{1}c_{1}c^{\dag}_{\bar{1}}c_{\bar{1}}\ket{\Psi}
&=& U( P_3 (\overline{u1}, \overline{u2},\overline{d1})t+ P_3 (\overline{u1}, \overline{u2},\overline{d2})t\nonumber\\
&& + P_5 ( \overline{g}, \overline{G}, \overline{u1}, \overline{u2}, \overline{d1})t^3+ P_5 ( \overline{g}, \overline{G}, \overline{u1}, \overline{u2}, \overline{d2})t^3 \nonumber\\
&&+  P_5 (\overline{g^{\dagger}}, \overline{G^{\dagger}}, \overline{u1}, \overline{u2}, \overline{d1})t^3 +  P_5 (\overline{g^{\dagger}}, \overline{G^{\dagger}}, \overline{u1}, \overline{u2}, \overline{d2})t^3 \nonumber\\
&& +P_7 (\overline{g^{\dagger}}, \overline{G^{\dagger}}, \overline{g}, \overline{G}, \overline{u1}, \overline{u2}, \overline{d1})t^5+P_7 (\overline{g^{\dagger}}, \overline{G^{\dagger}}, \overline{g}, \overline{G}, \overline{u1}, \overline{u2}, \overline{d2})t^5) \braket{tF|tF}\nonumber\\
\label{a7}
\end{eqnarray}
\begin{eqnarray}
\frac{\partial}{\partial G_{ef}} \sum_{<i,j>}t_{0ij}\bra{\Psi} c^{\dag}_{i}c_{j}\ket{\Psi} 
&=& 2( P_3 ( \overline{t}, \overline{g}, \overline{d1})t^2 -P_3(\overline{t}, \overline{g^{\dagger}}, \overline{d2})t^2 \nonumber\\
&&+  P_5(\overline{t}, \overline{g^{\dagger}}, \overline{G^{\dagger}}, \overline{g}, \overline{d1})t^4-  P_5(\overline{t}, \overline{g^{\dagger}}, \overline{g}, \overline{G}, \overline{d2})t^4) \braket{tF|tF}  \nonumber\\
\label{ap}
\end{eqnarray}
\begin{eqnarray}
\frac{\partial}{\partial G_{ef}}U\bra{\Psi} c^{\dag}_{1}c_{1}c^{\dag}_{\bar{1}}c_{\bar{1}}\ket{\Psi}
&=& 2U( P_4 ( \overline{g},\overline{u1}, \overline{u2}, \overline{d1})t^2 -P_4(\overline{g^{\dagger}},\overline{u1}, \overline{u2}, \overline{d2})t^2 \nonumber\\
&&+  P_6(\overline{g^{\dagger}}, \overline{G^{\dagger}}, \overline{g},\overline{u1}, \overline{u2}, \overline{d1})t^4-  P_6(\overline{g^{\dagger}}, \overline{g}, \overline{G}, \overline{u1}, \overline{u2},\overline{d2})t^4) \braket{tF|tF} \nonumber\\
\label{a9}
\end{eqnarray}

\section{Convergence of the AGP-CI calculation}
\label{app:decay_mode}
In AGP-CI calculation, we have done more than $50$ trial runs from different initial parameters to try to achieve full convergence, but the calculation with the standard conjugate gradient method was still trapped at a local minimum. This fact is reflected in the small but appreciable fluctuation of the calculated total energy as shown in Fig.\ \ref{fig:zua}. The dependence on the length of the Hubbard ring $N$ and the value of $U$ is not very smooth. However, the AGP-CI calculation was done only for comparison in this paper. So we used the value obtained here in Fig.\ \ref{fig:zu3}.

\begin{figure}[htbp]
  \begin{center}
    \begin{tabular}{c}
      \begin{minipage}{0.33\hsize}
        \begin{center}
        \hspace*{-3.5cm} 
          \includegraphics[clip, width=9.0cm]{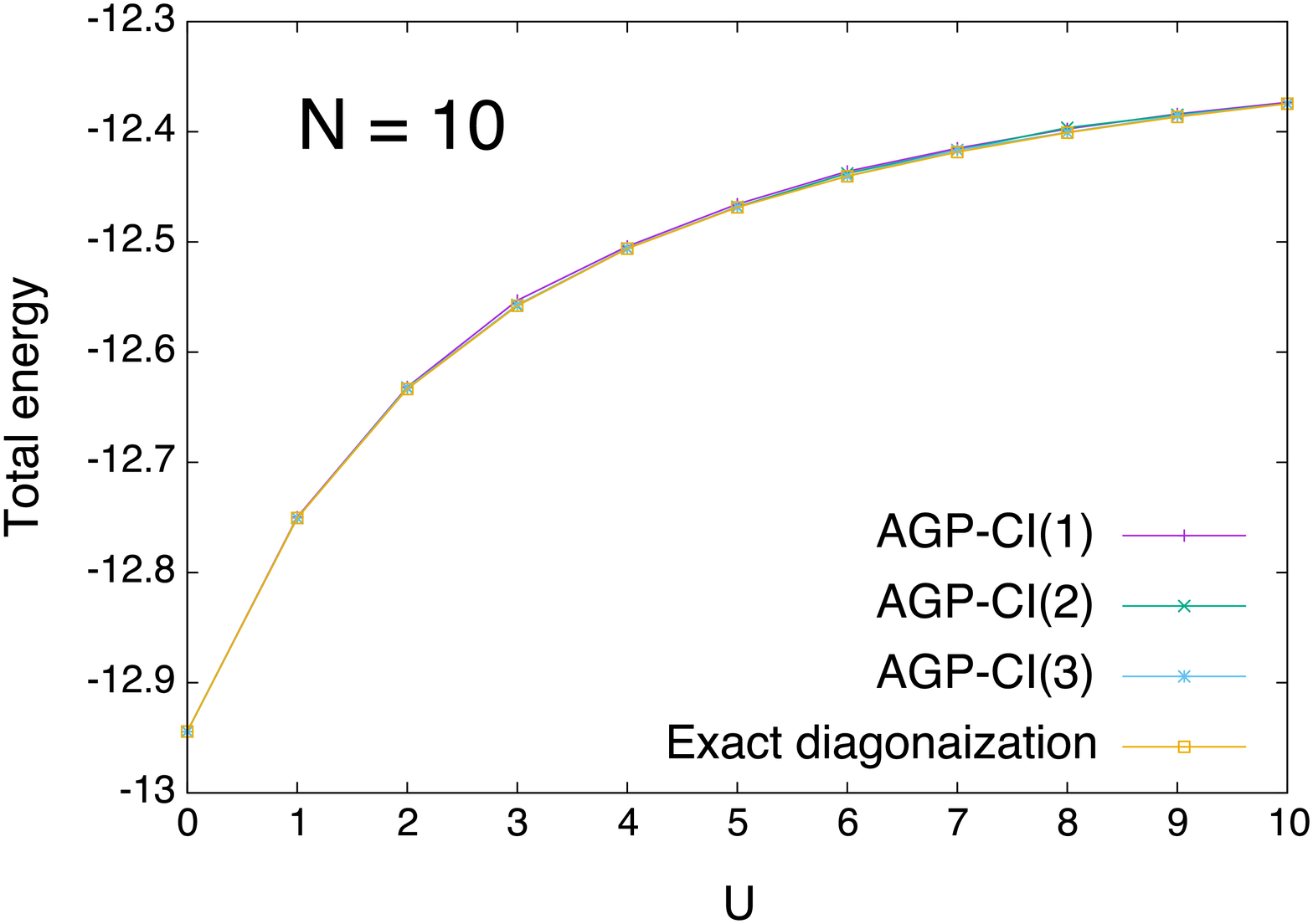}
          \hspace{1.6cm} 
        \end{center}
      \end{minipage}
      \begin{minipage}{0.33\hsize}
        \begin{center}
          \includegraphics[clip, width=9.0cm]{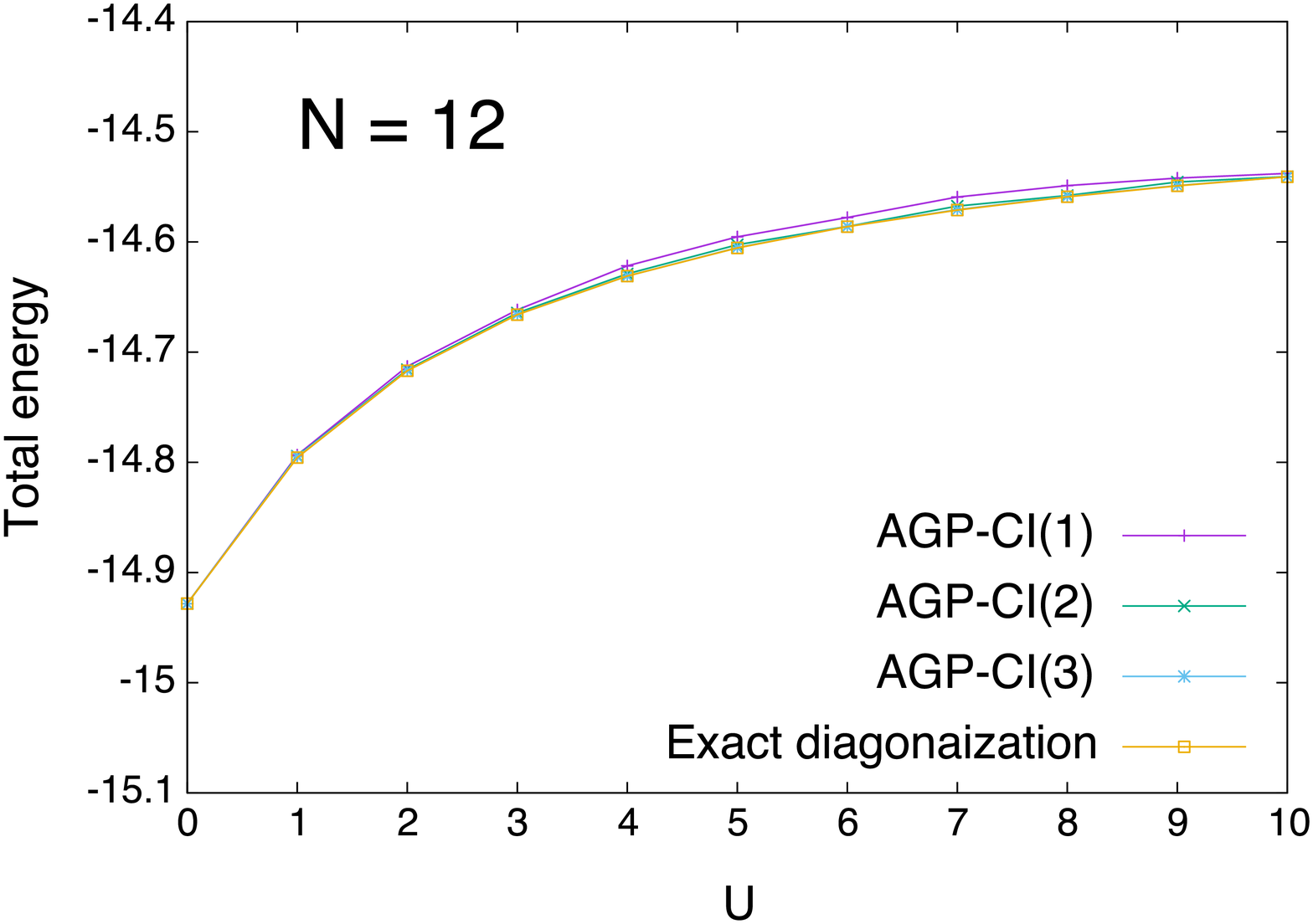}
          \hspace{1.6cm} 
        \end{center}
      \end{minipage}\\
      \begin{minipage}{0.33\hsize}
        \begin{center}
        \hspace*{-3.5cm} 
          \includegraphics[clip, width=9.0cm]{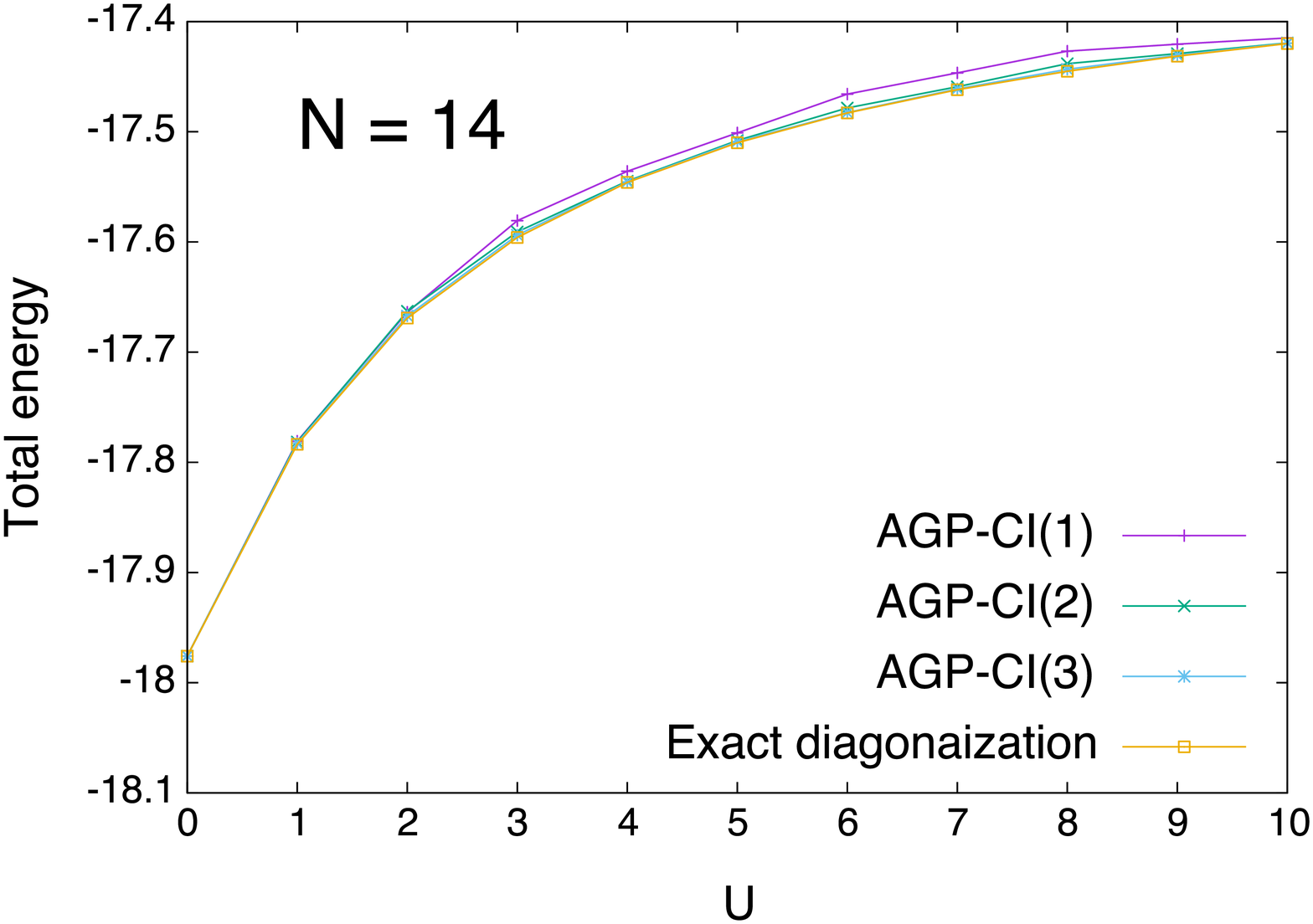}
          \hspace{1.6cm} 
        \end{center}
      \end{minipage} 
      \begin{minipage}{0.33\hsize}
        \begin{center}
          \includegraphics[clip, width=9.0cm]{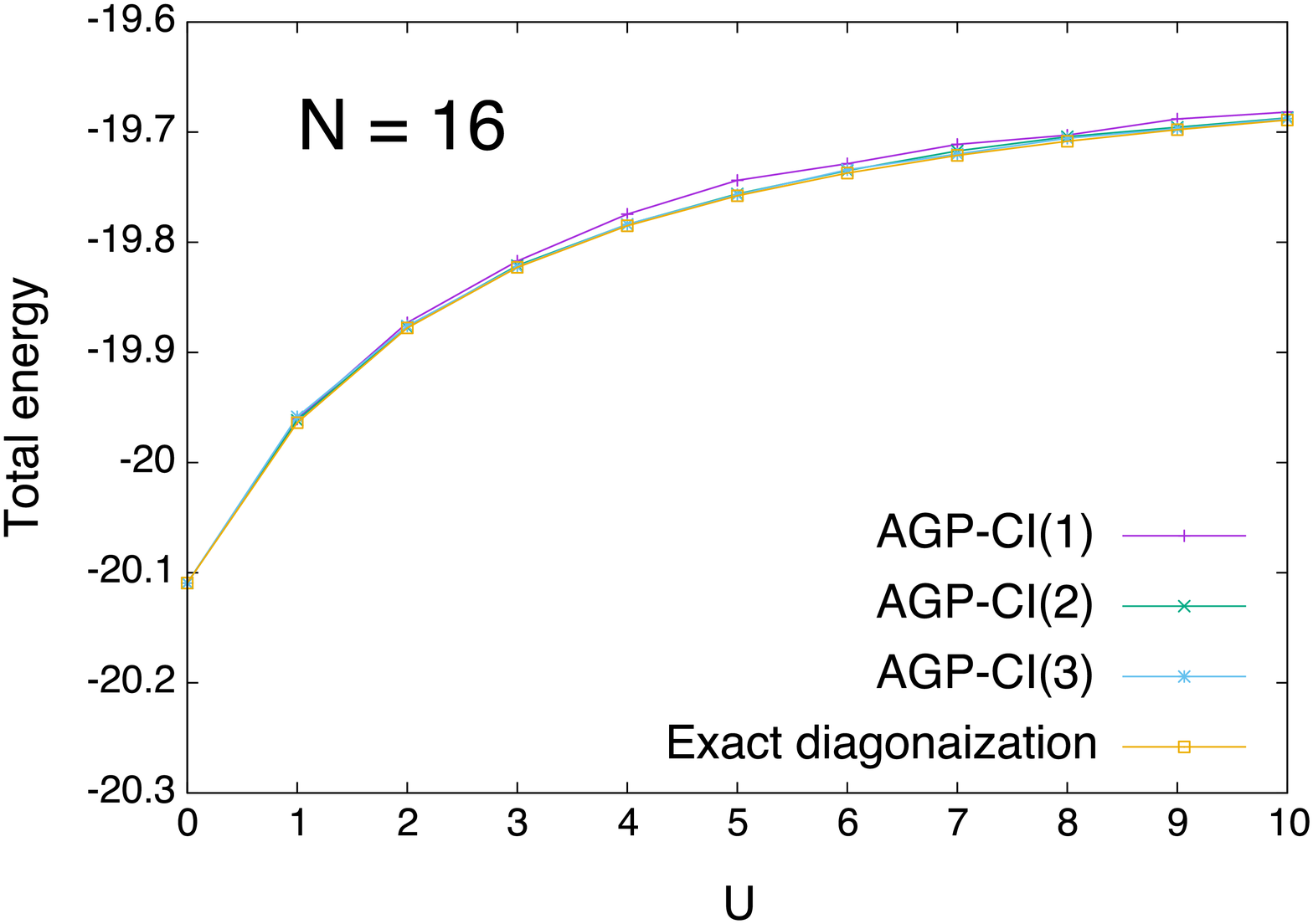}
          \hspace{1.6cm} 
        \end{center}
      \end{minipage}\\
       \begin{minipage}{0.33\hsize}
        \begin{center}
        \hspace*{-3.5cm} 
          \includegraphics[clip, width=9.0cm]{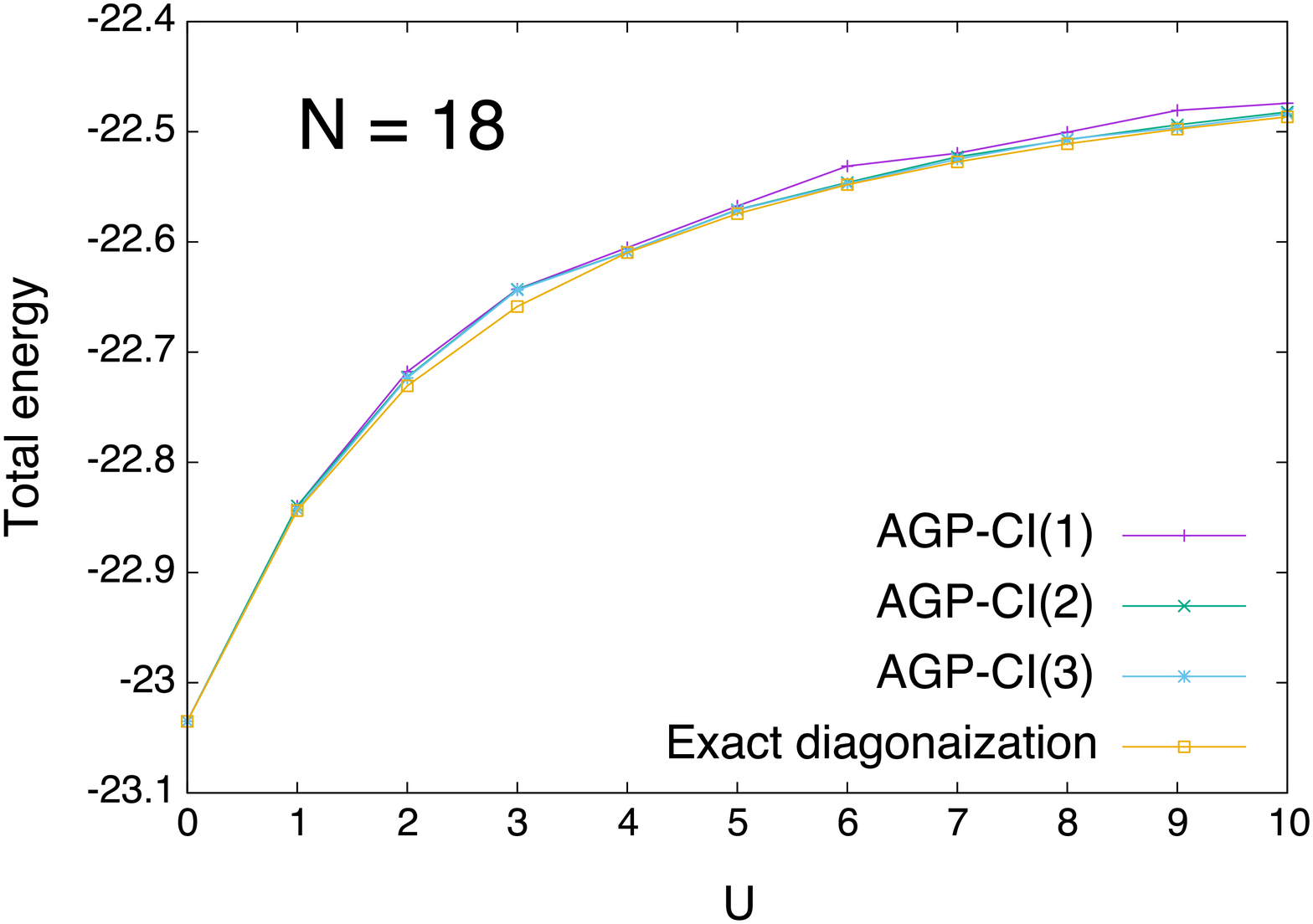}
          \hspace{1.6cm} 
        \end{center}
      \end{minipage}
       \begin{minipage}{0.33\hsize}
        \begin{center}
          \includegraphics[clip, width=9.0cm]{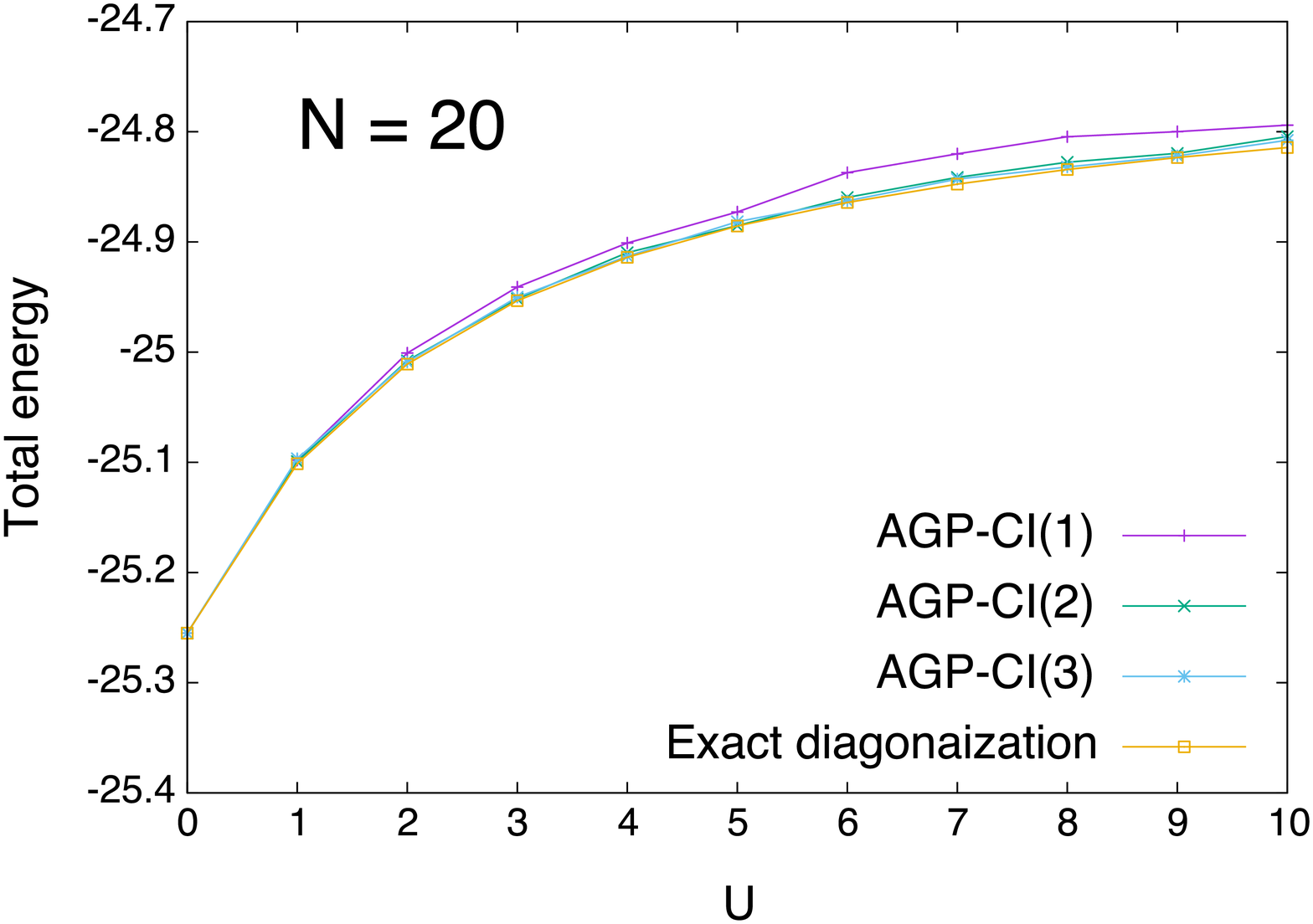}
          \hspace{1.6cm} 
        \end{center}
      \end{minipage}
     \end{tabular}
    \caption{(Color online) The AGP-CI total energy. The energies for the Hubbard ring of length $N=10$ to $N=20$ are shown for $U=0$ to $U=10$. The energy obtained by the exact diagonalization is shown for comparison.  }
    \label{fig:zua}
  \end{center}
\end{figure}

\end{widetext}

\bibliography{cite}



\end{document}